\documentstyle[12pt]{article}

\advance\textheight by \topskip
\newcommand{\sla}{\kern -5.4pt /}
\newcommand{\slaar}{\kern -7. pt / \kern 3.pt}
\newcommand{\Dir}{\kern -6.4pt\Big{/}}
\newcommand{\Dirin}{\kern -10.4pt\Big{/}\kern 4.4pt}
\newcommand{\DDir}{\kern -7.6pt\Big{/}}
\newcommand{\DGir}{\kern -6.0pt\Big{/}}

\newcommand{\be}{\begin{equation}}
\newcommand{\ee}{\end{equation}}
\newcommand{\bea}{\begin{eqnarray}}
\newcommand{\eea}{\end{eqnarray}}
\newcommand{\beanon}{\begin{eqnarray*}}
\newcommand{\eeanon}{\end{eqnarray*}}
\newcommand{\ba}{\begin{array}}
\newcommand{\ea}{\end{array}}
\newcommand{\bd}{\begin{description}}
\newcommand{\ed}{\end{description}}
\newcommand{\bi}{\begin{itemize}}
\newcommand{\ei}{\end{itemize}}
\newcommand{\ben}{\begin{enumerate}}
\newcommand{\een}{\end{enumerate}}
\newcommand{\bc}{\begin{center}}
\newcommand{\ec}{\end{center}}

\newcommand{\ar}{\rightarrow}

\newcommand{\vsk}{\vskip 10 pt\noindent}

\def\wm{M_{_W}}
\def\zm{M_{_Z}}
\def\gf{G_{\mu}}
\def\pd{{\it production $\times$ decay\ }}
\def\epem{$e^+ e^-$\ }

\def\app #1 #2 #3 {{\it  Acta Phys.Polon.} {#1} (#2) #3\ }
\def\ap #1 #2 #3 {{\it Ann. Phys. }{ #1} (#2) #3\ }
\def\intj #1 #2 #3{{\it Int. J. Mod. Phys.} {#1} (#2) #3\ }
\def\hpa #1 #2 #3{{\it Helv. Phys. Acta. }{ #1} #2) #3\ }
\def\mpl #1 #2 #3 {{\it Mod.~Phys.~Lett.} {#1} (#2) #3\ }
\def\np #1 #2 #3 {{\it Nucl.~Phys.} {#1} (#2) #3\ }
\def\pl #1 #2 #3 {{\it Phys.~Lett.} {#1} (#2) #3\ }
\def\pr #1 #2 #3 {{\it Phys.~Rev.} {#1} (#2) #3\ }
\def\prep #1 #2 #3 {{\it Phys.~Rep.} {#1} (#2) #3\ }
\def\prl #1 #2 #3 {{\it Phys.~Rev.~Lett.} {#1} (#2) #3\ }
\def\rmp #1 #2 #3 {{\it Rev. Mod. Phys.} {#1} (#2) #3\ }
\def\zp #1 #2 #3 {{\it Z.~Phys.} {#1} (#2) #3\ }
\def\epj #1 #2 #3 {{\it Eur.~Phys.~J.} {#1} (#2) #3\ }
\def\cpc #1 #2 #3 {{\it Comp. Phys. Commun.} {#1} (#2) #3\ }
\def\xx #1 #2 #3 {{#1}, (#2) #3\ }

\begin{document}
\tolerance=100000
\thispagestyle{empty}
\setcounter{page}{0}

\begin{flushright}
{\large DFTT 40/98}\\
{\large CTP-TAMU 31/98}\\
{\rm July 1998\hspace*{.5 truecm}}\\
\end{flushright}

\vspace*{\fill}

{\Large \bf \noindent
Higgs production in charged current six fermion processes
        at future $e^+e^-$ colliders.
\footnote{ Work supported in part by Ministero
dell' Universit\`a e della Ricerca Scientifica and\\[2 mm]
by National Science Foundation Grant No. PHY-9722090 (TAMU).\\[2 mm]
e-mail: accomando@chaos.tamu.edu, ballestrero@to.infn.it}}
\\[2.cm]
\bc
{\large Elena Accomando$^a$, Alessandro Ballestrero$^b$ 
   and Marco Pizzio$^b$}\\[.3 cm]

{\it $^a$Phys. Dept. Texas A\&M University\\ College Station TX 77843, USA}\\
\vsk{\it $^b$I.N.F.N., Sezione di Torino and \\
 Dipartimento di Fisica Teorica, Universit\`a di Torino}\\
{\it v. Giuria 1, 10125 Torino, Italy.}
\ec

\vspace*{\fill}

\begin{abstract}
{\normalsize\noindent
We study higgs  physics at future $e^+e^-$ colliders taking into account the 
full set of Feynman diagrams for six fermion final states, which are 
produced for higgs masses near or above  the two $W$'s threshold.
In particular we examine events where 
one isolated lepton or two isolated leptons of different flavours 
signal the presence of two $\rm W^*$'s. For these charged current processes,
a detailed analysis of  the 
relevance of  irreducible and QCD background
shows that  appropriate cuts are generally sufficient to deal with it 
in the case of reconstructed
or missing higgs mass distributions. These latter are  however affected 
by a non negligible distortion and shift of the maximum.   
} 
\end{abstract}

\vspace*{\fill}

\begin{flushleft}
\end{flushleft}

\newpage

\section{ Introduction }
One of the most challenging present physical issues is  the search of the 
higgs particle and the detailed study of its properties. 
Precision data fits\cite{pc} tend to exclude a higgs heavier than about 
300 GeV, and favour a lighter mass. The experimental lower limit for a 
Standard Model higgs is now of the order of 90 GeV. For the Minimal
Supersymmetric Standard Model it is
about 10 GeV lower, while theoretical arguments put an upper limit at around
130 GeV\cite{lim}.
 If the higgs is not discovered at LEP2 or Tevatron,
 its mass will most likely turn out to be higher than 130 GeV. 
In such a case LHC will probably find it, and  the
future \epem Linear Collider will be an ideal machine
to study its properties, 
being complementary to the LHC in this respect.
\par
In this paper we analyze Standard Model higgs physics at 
 the Linear Collider 
supposing that its mass is heavier than about 140 GeV, in the so called
intermediate mass region.
As it is well known, in such a case  the higgs
 mainly decays into two (possibly
virtual) $W$'s and the most important production channels are 
$hZ$ production (up to 500 GeV) and  $WW$ fusion, which dominates
at higher energies. The final states to be studied correspond therefore
to six fermions, four from the $h\ar W^*W^* \ar 4f$ decay and two either from
the $Z$ decay or the $\nu_e$'s in  the
$WW$ fusion. 

Six fermion final states  receive contributions from a great 
number
of different Feynman diagrams. Only few of them correspond to higgs production
and decay. All the others come from  resonant processes
(like $t\bar t$ or $WWZ$ production), partial resonant  (as for instance
$\nu_e \bar \nu_e WW$) or non resonant ones. 
All these diagrams, whose number is of the order of several
hundreds, correspond to the same final state, interfere among
themselves and are to be considered for a complete calculation. Thus, only
appropriate cuts can   enhance
the signal contributions and separate
them from the  irreducible background which comes from all other diagrams.
It should be noticed that different resonant contributions might be considered
either as signal or as irreducible background depending on the 
process under study and their interplay can be consistently analyzed 
only by complete calculations.  

Intermediate mass higgs studies at  the
Linear Collider have been performed
mainly in the \pd approximation (see for instance ref.\cite{pd}), 
where radiative corrections to both production and decay are available 
\cite{rc}. 
The alternative approach of considering all contributions to a more complete 
final state as $l\bar l WW$, $\nu\bar\nu WW$ or $b \bar b WW$ has been considered a 
few years ago\cite{bbww}. 
 Recently three groups\cite{noi}\cite{kuri}\cite{pv} have 
started computing full tree level cross sections for six fermion final states 
at the Linear Collider. 
These  computations have been applied to study the phenomenology of 
$t\bar t$\cite{noi}\cite{kuri}, $WWZ$\cite{noi} and 
higgs\cite{pv}\cite{noi2} production. 
In ref.\cite{pv} a detailed analysis
of  $e^+e^-\ar\mu^+ \mu^- l\bar\nu_l u \bar d$ $(l=e,\tau)$ 
and of all reactions of the type $e^+e^-\ar l^+l^-\nu\bar\nu q\bar q$
($q\neq b$) has been performed in connection with higgs physics. 
In  ref.~\cite{noi2} some partial results of the present studies have been 
summarized. 

In the following  we  systematically study and discuss all relevant 
6f processes in which one or more isolated leptons of different
flavours make it possible to 
identify a final state corresponding  to the production of two $W$'s.
The reason of our choice is simple: these charged current processes constitute
about 52\% of the total $hZ\ar WWZ$ signal and they are the cleanest processes
both from the phenomenological and experimental point of view.
In fact, just because of their final state, they allow to get rid of 
almost all background corresponding to 
processes in which two  $W$'s have not been formed (the only relevant exception
corresponds to  $l \bar \nu q\bar q' g g$ which we will examine in detail). 
For this reason, it is natural to start with studying irreducible 
backgrounds and realistic mass distributions without the additional 
complications arising for instance from six jets QCD, ZZZ, or four fermion 
processes. 
So we will not consider in this paper six quark final states, which represent
about one third of $hZ\ar WWZ$ signal but are obviously affected by a severe 
six jet QCD background. 
 We will also not discuss the properties of
 six  fermion final states which can come from three $Z$'s decay, as it happens
for $l\bar l \nu \bar \nu  q \bar q $, for 4 quarks with a couple 
$l\bar l$ of same flavour leptons, and for 4 $q$'s + 2 $\nu$'s.  
Out of these last three processes the first, which corresponds to  about 
2.6\%  
of $hZ\ar WWZ$  signal and may also be affected  by 
$t\bar t$ background, has already been discussed in the literature \cite{pv}.
The other two represent respectively around 4.4\% and 8.8\% of 
$hZ\ar WWZ$ signal and their $ZZZ$ and QCD backgrounds have to be carefully
analyzed. In the case of  4 $q$'s + 2 $\nu$'s
and $l\bar l \nu \bar \nu  q \bar q $, also  
four fermion processes can be dangerous and must be considered.

Being  interested in  higgs signal, one has of course always to deal with 
$WWZ$ background, but this is  really irreducible in the sense 
that it cannot be eliminated just by choosing a particular set of final states.
 A careful
study of distributions and of possible cuts is therefore  unavoidable.

The plan of the paper is the following. In sect.~2 we give some detail
of the processes and of the method used for their calculation. In section ~3 
we examine the results for final states with 2 quarks and in section~4 those
with four quarks. We draw some conclusions in section~5.

\section{Processes and their calculation}

As already mentioned in the introduction, we will 
consider
in detail  
processes with one isolated lepton
like $l\: \nu_l\: +\: 4\: q's$ or $l\: \nu_l\: +\: l'\: \bar l'\: +\: 2\: q's$,
and
those
 with missing energy and two leptons of different flavour and charge   
like $l\: \nu_l\: +\: l'\: \nu_{l'}\: +\: 2\: q's$. 
We use $l\: \nu_l$ to indicate both $l^-\: \bar \nu_l$ and $l^+\: \nu_l$.
The first two cases  represent respectively about 31\% and 4.4\% of $hZ\ar WWZ$ 
signal, while the third one amounts to about 4.1\% ($q\neq b$),  as one 
simply deduces from $W$ and $Z$ branching ratios. 
Let us notice that for the first two final states, the 
isolated lepton plus missing energy indicates that most likely two 
$W's$ have been formed in the 
process. The same indication is given in 
the third  process
by the two isolated leptons of different flavour and charge, which 
imply the presence of their corresponding neutrinos. 
We will not explicitly discuss the final state
$l\: \nu_l$ + $l'$ $\nu_{l'}$ + $l''\; \bar l''$, as  it amounts only to 
.7\% of the $hZ\ar WWZ$ events and it is very 
similar to $l\: \nu_l\: +\: l'\: \bar l'\: +\:2\: q's$.

All calculations have been performed with the program SIXPHACT, which has
already been used for $t\bar t$ and $WWZ$ studies\cite{noi}. 
In the present version we have added 
the diagrams containing higgs exchanges and
other mappings for the phase space, in order to account for the resonant 
peaking structures
corresponding to higgs events. A detailed description of how the 
computation is organized can be found in ref.~\cite{noi}. 
At present the program can compute all charged current (CC) 6f final states,
i.e. those final states which can derive from the decay of two $W$'s and a $Z$,
but cannot reconstruct the ones coming from three $Z$'s.
  All tree level Feynman diagrams are exactly evaluated in an helicity
amplitude formalism\cite{method}. Their number is  at least 
of the order of 200 when no exchanges among identical particles
are possible.  If we consider for instance final states with one isolated 
lepton  and four quarks,  $\mu \bar \nu u \bar d b \bar b$ has 232 diagrams 
 (23 of which are due to higgs) and $e\bar \nu u \bar d u \bar u$ 840 (4 with 
  higgs).

SIXPHACT  can account for  initial state radiation (ISR),
  beamstrahlung (BMS) with  a link to CIRCE\cite{circe},
  naive QCD corrections (exact in \pd no cuts limit). It has
  exact $b$ and $t$ fermion masses both in the amplitudes and in the phase
  space. All results given in the following have
  been   computed with ISR, BMS, NQCD. CPU times depend of course on 
 the particular final processes, on the cuts,
and the options required. As an indication, a process like $e^+e^-\ar\mu^+
\mu^- e \bar\nu_e u \bar d$ (418 diagrams) 
with ISR, BMS and NQCD takes about 
20 minutes for 3 per mille, 3 hours for per mille  and
one day for .2 per mille accuracy on a DEC ALPHA station. 

For the numerical part we have used  
$s_{_W}^2 = 1 - {{\wm^2}/{\zm^2}}, \quad
g^2 = 4{\sqrt 2}\gf\wm^2$
and the  input masses  $m_Z=91.1888$ GeV, $m_W=80.23$ GeV. We have chosen
 $m_t=175$ GeV and for $m_b$ the running mass value $m_b=2.7$ GeV.
For the strong corrections to $Z$, $W$, $t$ and $h$ decay widths
and vertices  we have used
$\alpha_s(m_Z)=$ 0.123 and let it evolve to the appropriate scales.

We have  implemented the following general set of cuts, appropriate for
 the Linear Collider studies:
\bc
 jet(quark) energy $> 3$ GeV;\quad
 lepton energy $> 1$ GeV;\quad
 jet-jet mass $> 10$~GeV;\\
 lepton-beam angle $> 10^{\circ}$;\quad
 jet-beam angle $> 5^{\circ}$; \quad
 lepton-jet angle $> 5^{\circ}$.
\ec
For the different processes we have sometimes used  other cuts which 
will  be described in due time.

In the following we will mainly study invariant mass distributions,  
from which
one can hope to extract informations on higgs mass and width  and  study 
appropriate cuts. An example of these
distributions is given in fig.~1 for the
process $\mu \bar \nu u \bar d b \bar b$. Here and in the following
figures, we do not report the errors on the single bins as the statistical
precision is such that they are practically not distinguishable on the plots.
The distributions in fig.~1 are obtained
with signal diagrams only. In practice, apart from the off-shellness
of higgs, $W$'s and $Z$ and from spin correlations,
 these distributions correspond
to the ones one would obtain in the \pd approximation. 
In the upper part of the figure we have used a logarithmic scale.
The three curves correspond to the exact higgs (i.e. $\mu \bar \nu u \bar d$)
distribution (continuous line), to the reconstructed mass (dashed) and
to the missing one (chaindot). As it is obvious, the exact $\mu \bar \nu u
\bar d$ invariant mass distribution cannot be experimentally determined, due to 
the fact that the
neutrino four momentum cannot be measured. One can try to reconstruct
it from the missing momentum and energy. But the presence of ISR and BMS, which
also contribute to the missing momentum and energy, 
introduces  an error in this determination. The
best one can do is to ascribe
all missing 3-momentum to the neutrino and determine its energy from
 the on shell condition. We name reconstructed mass the one
obtained by using 
 this approximately reconstructed 4-momentum.
The missing distribution corresponds instead to the invariant mass of the 
4-momentum recoiling against the 
particles ($b$'s in this case) decaying from the $Z$. In absence of ISR and BMS 
this would be the exact $\mu \bar \nu u \bar d$ invariant mass. 
Fig.~1 shows clearly the differences between the two approximated, more
realistic, distributions and the exact one. In particular, in the lower part
 of the figure  the difference between reconstructed and missing mass 
is reported in a  linear scale. 
 Of the two, the latter is much more distorted.
This feature is not peculiar of the specific final state considered.
It is therefore evident that, whenever possible, the reconstructed
mass will be measured and studied. 

\section{ Final states with two quarks.}
In this section we will examine processes of the type
 $l\: \nu_l\: +\: l'\: \bar l'\: +\: 2\: q's$
and 
$l\: \nu_l\: +\: l'\: \nu_{l'}\: +\: 2\: q's \: (q\neq b)$. 
Even if the two final states together only amount to
about 10\% of the 
$hZ\ar WWZ$ 
signal, they are 
a priori 
particularly interesting as they  have neither 
QCD nor $t\bar t$ background. With $q$ we denote here any
light quark $u,d,c,s$,  so one can actually measure the second reaction
only if one excludes the $b$'s with b-tagging. Otherwise one has to face
the huge $t\bar t$ background. 

For the first type reaction we consider 
$e\bar \nu u \bar d \mu^+\mu^-$. At 500 GeV the signal cross section for
$m_h=200$~GeV is 0.03501(2) fb, while the total one is 0.07824(3) fb. 
As usual we indicate  in brackets the
error on the last digit and we have applied two additional cuts:
$|m(\mu^+\mu^-)-m_Z|<20$ GeV and $|m(u\bar d)-m_W|<20$ GeV.
The two cross sections above show that the irreducible background given 
by all non higgs diagrams  can often be of the same order or even bigger
 than the signal. 
In practice however one does not have to worry about this fact. 
 This is evident  from fig.~2, where the
reconstructed mass distribution is
reported for the total cross sections (for $m_h=150,170,200,250$ GeV) and the 
background only (continuous line). 
 One can see that the part of the background under the higgs mass peaks 
is almost irrelevant. In fact, considering an interval of 20 GeV around
the peak,   the ratio SB between signal and background    for 
$m_h=$200 GeV is 10.0 at 500 GeV and 14.0 at 360 GeV. In the worst case 
reported in the figure, namely at $m_h=$250 GeV, SB is 5.1 at 500 GeV 
and 3.7 at 360 GeV.
As it is obvious the background in this case receives its greatest contribution
from $WWZ$ resonant diagrams. The fact that it is not important when
considering invariant mass distributions is of course common to all processes
and not specific to the one at hand.
Another general feature that can be easily seen in fig.~2 is that
 the signal for $m_h=$150 GeV, which is below $WW$ threshold, is clearly
visible. This happens even if we have applied a cut on $m(u\bar d)$ in order 
to constrain it  to be in the vicinity of the $W$ mass.

It is interesting to evaluate 
 also the signal to square root of background ratio SSB. We do this
always by considering an interval of 20 GeV
around the nominal higgs mass, by assuming a luminosity of $50 fb^{-1}$ and
multiplying both signal and background by  8 to approximately account
for the multiplicity of the analogous channels. For 
 $l\: \nu_l\: +\: l'\: \bar l'\: +\: 2\: q's$
and 
$l\: \nu_l\: +\: l'\: \nu_{l'}\: +\: 2\: q's \: (q\neq b)$, a 
 factor $\approx$ 4  accounts for the different $q$'s 
and a factor $\approx$ 2 for choosing $l=e$ and $l'=\mu$ or viceversa.
Of course the SSB numbers improve by more than a factor 3 for 
the higher luminosity option of  $500 fb^{-1}$, and this might be important
for high higgs masses. 
For $m_h=$200 GeV  SSB is 10 at 500 GeV and 15 at 360 GeV; at $m_h=$250 GeV, 
SSB is 5.7 at 500 GeV and 4.1 at 360 GeV.

As far as  $l\: \nu_l\: +\: l'\: \nu_{l'}\: +\: 2\: q's$ is concerned, let us
consider specifically $\mu\bar \nu e^+\nu s \bar s$. For this reaction we
apply only the additional cut $|m(s\bar s)-m_Z|<20$.
Also in this case if we consider the total cross section and the
signal one we find that the background can be of the same order of the signal.
For instance at 500 GeV for $m_h=170$~GeV the total cross section is 
0.14498(3) fb and the signal amounts to 0.08077(1) fb.  So we must consider the 
invariant mass distributions. 
In this case  the reconstructed mass is not experimentally viable, so one
looks at the missing one. The distributions are reported in fig.~3. Once again
one can compute SB and SSB for $m_h=$200 and 250 GeV. At lower $m_h$ the 
background is less and less  relevant. In the first case we have 
SB=7.9, SSB=8.5 at 500 GeV and SB=13.5, SSB=16.5 at 360 GeV. 
 In the latter, one finds respectively SB=3.6, SSB=4.8 and SB=3.1, SSB=4.5.
It has however to be noticed that  the signal missing mass distributions are
very asymmetric and their form is  rather unrealistic. We have therefore
examined the case in which  the mass reconstruction is affected by a  gaussian 
"experimental error" of 5 GeV. This means that every missing mass computed
by our program has been displaced with a gaussian probability whose
standard deviation is 5 GeV.  With this we try to simulate in a very
simple and qualitative way the uncertainties introduced by the hadronization 
and the  reconstruction from real particles of the invariant mass.
The result is reported in fig.~4. One may notice that the 
smearing 
makes the distributions   more
symmetric, but it produces a
sizeable shift of the maximum of the order of a few GeV. This is
a feature that has to be carefully taken into account when one
wants to study  higgs properties in processes with two neutrinos or in general
using missing mass distributions.
As a consequence of these results and since the branching
ratio of this channel is not more than 5\%, our conclusion is that
$l\: \nu_l\: +\: l'\: \nu_{l'}\: +\: 2\: q's$ is not a very good candidate
for higgs studies.  Including all possible  final states with
two $\nu$'s and 2 or 4 leptons,  the total branching ratio  increases up to 
about 8.8\%,
but in that case  the huge background from ZZ into four fermions could also
affect the final state with two charged leptons of the same flavour.
 
\section{ Final states with four quarks and one isolated lepton.}

Final states with four quarks and one isolated lepton correspond to  about 
one third  of higgs $hZ\ar WWZ$ events. They are however affected by 
$t\bar t$ and QCD  background.  
The latter may come  from both 2 gluons + 2 quarks + $l\nu$ events and  
from the contribution to the usual  $l \nu + 4 q's$ final state 
due to diagrams
with gluon exchange between the two quark lines.

One may consider  processes with or without 2 $b$'s in the final state.
They can be distinguished with b-tagging, although one has to take into account
that such an identification has always 
 some degree of error. If two $b$'s 
and one isolated lepton are present, the event cannot correspond to 
2 gluons + 2 quarks + $l\nu$ production, and one does not have to worry 
about  such  background.
Moreover if a $Z$ has been produced,  the two $b$'s are necessarily 
 its decay products.
So one can apply a cut, forcing the $b\bar b$ invariant mass to be in the
 vicinity of $M_Z$. This kind of
cut does not reduce the higgs signal,
while it strongly suppresses   QCD background coming from gluon exchange 
diagrams. 
With two $b$'s in the final states, the largest background is  
that due to $t\bar t$ production and
decay. The above mentioned cut is efficient also in this case, as one does not
 expect that the $b$'s coming from $t\bar t$ decay 
prefer an invariant mass near the $Z$. 
 However the $t\bar t$ cross section is about 10 times bigger
than the $hZ\ar W^+W^-b\bar b$ production.
Therefore one has to carefully compute the 
effects  of the  $b\bar b$ cut. In the upper part of fig.~5 we present 
the contribution of the background to the 
distributions of the true, reconstructed and missing invariant higgs mass for 
$\mu\bar \nu u\bar d b \bar b$, computed 
 by imposing that 
$|m(u\bar d)-m_W|<$20 GeV and  $|m(b\bar b)-m_Z|<$20 GeV. Only the latter
cut is relevant to the $t\bar t$ production cross section
 which is reduced by  a factor 5.  

If  one takes the other possibility and excludes  $b\bar b$ events, one 
does not 
have to worry about $t\bar t$ (apart from imperfect b-tagging), but QCD
and EW backgrounds have to be considered. 
As far as the signal is concerned, there are no important differences
with  respect to the $b\bar b$ case, apart for the fact that one 
can now sum over the different light flavours emitted by the $Z$. 
We do not report
 the plot equivalent to fig.~1 for the signal with $s\bar s$ instead of $b\bar
 b$, just because  the differences are at percent level
even if the cuts are now  different.
In fact having 4 light quarks in the
final state, it is not  known which jets are produced by the $Z$ decay  
and which
by the $W$. Therefore we  choose to accept the event if 
 at least one  pair of couples $q q'$, $q''q'''$ is such that $|m(q q')-m_Z|$
and $|m(q'' q''')-m_W|$ are less than 20 GeV.
This cut is applied also in the following to all results with four
light quarks in the final state. 

 In the lower part of fig.~5 one can
see the  background for  $\mu\bar\nu u\bar d s \bar s$, 
due to  EW diagrams and to diagrams with
gluon exchange. The curves marked QCD  refer only to this irreducible
 background and not to $qq+ gg + l\nu$. 
The process $\mu\bar\nu u\bar d s \bar s$ is obviously not 
experimentally distinguishable, but it has been
considered here because it is at all similar to 
$\mu\bar\nu u\bar d b \bar b$ without
single and double resonant top diagram contributions. 
The tiny differences due to  b mass are not relevant in this context. 
By comparing the upper and the lower part of fig.~5, one therefore concludes 
that the irreducible background coming  from $t\bar t$
production and decay
is  more important by a factor of order 30 than the remaining EW contributions.
The irreducible background coming from gluon exchange is completely irrelevant
as it is  about three  order of magnitudes  less  than the 
 $t\bar t$ one.
\par
In fig.~6 we study the importance of the various backgrounds when
there is no  $b\bar b$ pair in the final state.
In order to consider a realistic cross section we sum over all remaining 
$q\bar q$ flavours ($q= u, d, c, s$). The upper part of fig.~6 leads to the same
conclusions as the lower one of fig.~5 for the irreducible background via
gluon exchange: the curve marked $QCD \times 20$, which corresponds to 
this background
multiplied by a factor 20 is still lower than the EW one, thus
confirming its complete irrelevance.
In the lower part of fig.~6 we report the background coming from 
$\mu\bar\nu u \bar d g g $ final state. This turns out to be almost as important
as  the EW one. We remind that in this case the EW does not contain any
contribution from $t\bar t$ states. In the following we will 
not sum  the $l \bar \nu q \bar q' gg$ background to the irreducible 
EW one, because it is conceivable that it may be further reduced by
experimental cuts
or by the possibility to distinguish among quark and gluon jets.
 Moreover, it is immediate to evaluate it from 
fig.~6 and we will see in the following that the irreducible background not
coming from $t\bar t$  is of scarce significance even if one multiplies it by 
an approximate factor two, to account for two quark plus two gluon jets.

Coming to a more specific analysis of signal-background interplay,
let us  
consider
 the process $\mu \bar \nu u\bar d b\bar b$. 
Processes of this type with two $b$'s are  of the order of 6.5\% of 
total $hZ\ar WWZ$  signal. 
 In fig.~7 we show the reconstructed mass distribution 
for the background and for the complete cross sections 
corresponding to $m_h=$150, 170, 200, 250 GeV. 
One notices immediately in the lower part of the figure that 
up to 200 GeV the cut $|m(b\bar b)-m_Z|<$20 GeV is certainly sufficient.
For $m_h=200$ GeV the ratio SB is 1.4 and  
SSB is 8.1, while for $m_h=250$ GeV one gets  
SB=.2 and SSB=2.5.  In the latter case, the number of
signal events is 29.7 while the background events are 139.
For the processes in this section the factor 8, by which we have 
 multiplied signal and background,
approximately corresponds to the possibility of having isolated  electrons
or muons  of different charge and  $c\bar s$ in place of $u\bar d$. 
In the upper part of fig.~7 we compute the effects on the distribution
of an additional cut regarding the invariant mass of the top. 
 For top events, $ u \bar d b$  reconstructs the mass
of the top. As one cannot distinguish  $b$ from $\bar b$ experimentally,  
we  impose  that $|M-m_{top}|>$40 GeV both for $M=m(bu\bar d)$ and 
for $M=m(\bar bu\bar d)$.  From the comparison of the lower and upper part
of fig.~7,  one
deduces that  such a cut reduces the signal  by about a
factor 2.5, 
but SB and SSB become respectively 6.0 (2.8) and 10.2 (5.2) for 
$m_h=200(250)$ GeV,  still leaving 9.6 events when $m_h=250$ GeV.

If one excludes the $b$'s and considers only final states 
with light quarks, the
corresponding distribution is represented in the upper part of fig.~8 
where 
 we sum over the quark flavours as indicated in the plot, 
taking properly into account the additional contributions arising in the case 
when identical quarks are
present in the final state. 
One can see that the situation is very clean.
  The background from non perfect b-tagging  is not taken into account
 and depends on the probability that a
$b$ is misidentified. One can however have an idea
of its importance considering the extreme case in which there is no b-tagging
at all, as we do in the following. 

It is interesting to examine the case in which one simply sums over 
all quark flavours ($u,d,c,s,b$) for the  $q\bar q$ pairs,
since the higgs signal of course increases in this case.
The corresponding distribution is presented in
the lowest part of fig~8.
 In this case  also for the $b \bar b$ events the
cuts are the same as for light quarks: $|m(q q')-m_Z|<$20 GeV
and $|m(q'' q''')-m_W|<$20 GeV as explained before.
One immediately deduces from the figure that below $m_h\approx 180$~GeV the 
signal can be measured without any problem. But also for higher masses the
situation is not bad: even if at $m_h=$200 (250) GeV SB  is .7(.4),  SSB 
is 12.2(7.4). It would be possible, if necessary, to impose cuts on the 
invariant masses of three  jets, rejecting events
in which any jet triplet form an invariant mass in the vicinity of the 
top.

Finally in fig.~9 we have examined the possible effects on the distributions of
the experimental error in the determination of the reconstructed higgs mass.
 We have assumed a 5 GeV gaussian error as before. 
 One can see that the shape of the signal distributions
changes rather significantly, but as far as SB and SSB is concerned the 
difference is almost irrelevant: 
SB=.7(.4) SSB=11.6(6.8) at 200(250) GeV. 

\section{Conclusions}
By means of complete 6f calculations performed with SIXPHACT,
we have examined higgs processes with  one or two isolated 
leptons of different flavour
in the 
final state, in order
to eliminate as much as possible the background not containing two $W$'s. 
We have moreover applied cuts to eliminate $t\bar t$ and studied reconstructed
and missing invariant mass distributions at 500 and 360 GeV.

The main results may be summarized as follows.

For processes  $l\: \nu_l\: +\: l'\: \bar l'\: +\: 2\: q's$,
where QCD and $t\bar t$ background are absent, the signal is very clean  
and the background has hardly any effect around
the reconstructed higgs invariant mass. Even \pd approximation could be used
in that region, while for the total cross section the contribution of the
irreducible background can be of the same order of the signal.

The same applies to 
$l\: \nu_l\: +\: l'\: \nu_{l'}\: +\: 2\: q's \: (q\neq b)$, but in 
this case one has to
consider the missing energy which entails a substantial distortion and 
a shift of 
some GeV in the distribution. As b-tagging might moreover leave in practice
a part of $t\bar t$ background, it seems that this reaction is not the
best candidate for higgs studies.

The processes  $l\: \nu_l\: +\: 4\: q's$ with four quarks  correspond
to about one third of $hZ\ar WWZ$ signal and are therefore more important.
Irreducible  QCD background due to diagrams with gluon exchange 
is completely irrelevant. The one coming from $l\bar \nu q \bar q' gg$
is  of the same order of
the electroweak one and has little practical effect.
The main source of 
background comes from
$t\bar t$ production and decay. It turns out that it can  be kept under 
control,
at least for not too high higgs masses, both if one sums over all
 final state quarks or if one looks specifically for a $b\bar b$ couple in 
the final state.

\vfill\eject

\begin{figure}[h]
\vspace{1.cm}
\begin{center}
\unitlength 1cm
\begin{picture}(14.,15.)
\put(-1.8,-2.){\includegraphics{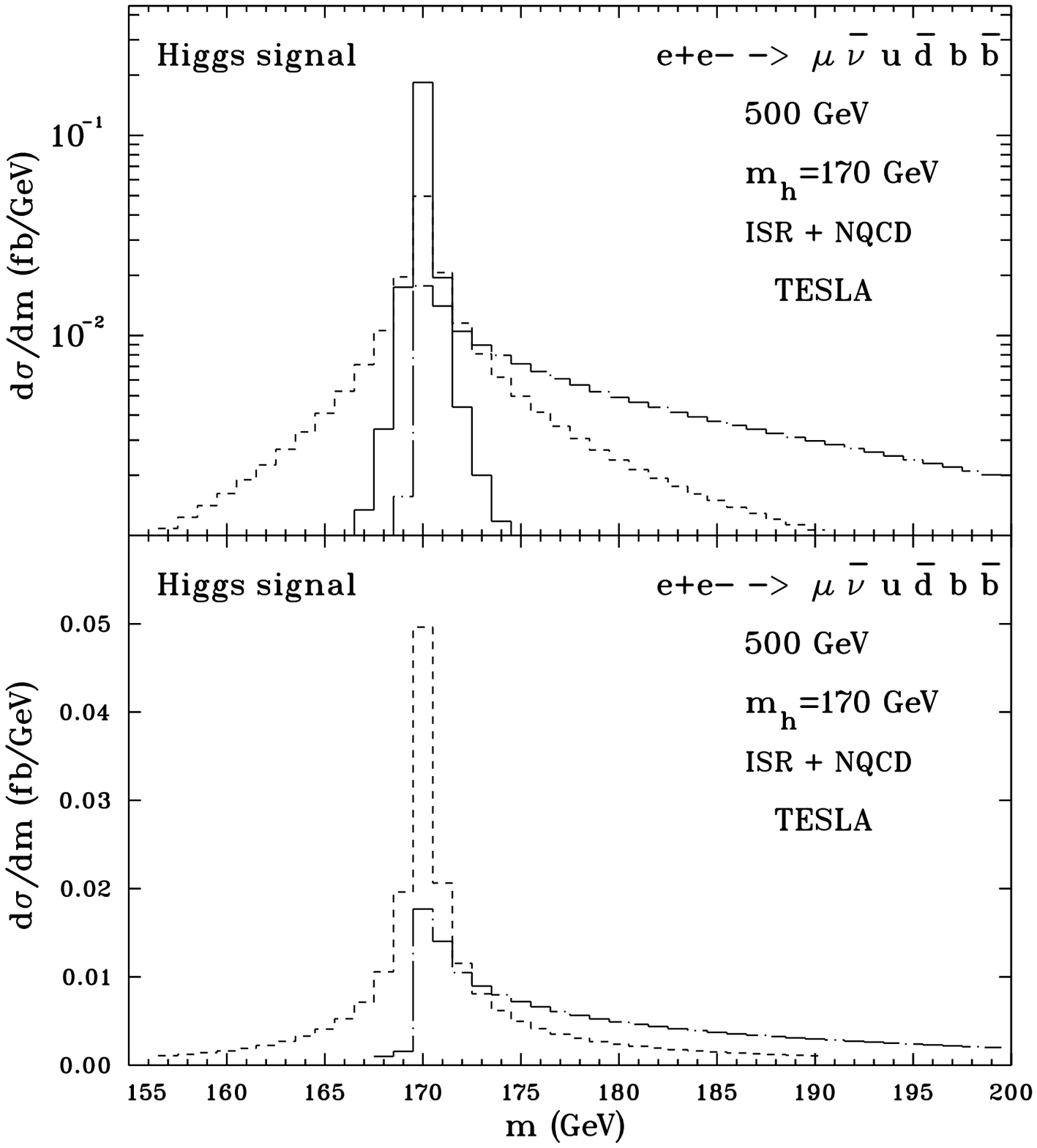}}
\end{picture}
\end{center}
\vspace{0.1cm}
\caption[]{Invariant mass distributions for higgs signal. The continuous line 
corresponds to ($\mu \bar \nu
u \bar d$)  mass, the dashed line to the {\it reconstructed} and the chaindot
 to the {\it missing} one.
Cuts: $|m(b \bar b)-m_Z|<20$ GeV, $|m(u\bar d)-m_W|<20$ GeV.
}
\label{h1}
\end{figure}

\begin{figure}[h]
\vspace{0.1cm}
\begin{center}
\unitlength 1cm
\begin{picture}(14.,15.)
\put(-1.8,-2.){\includegraphics{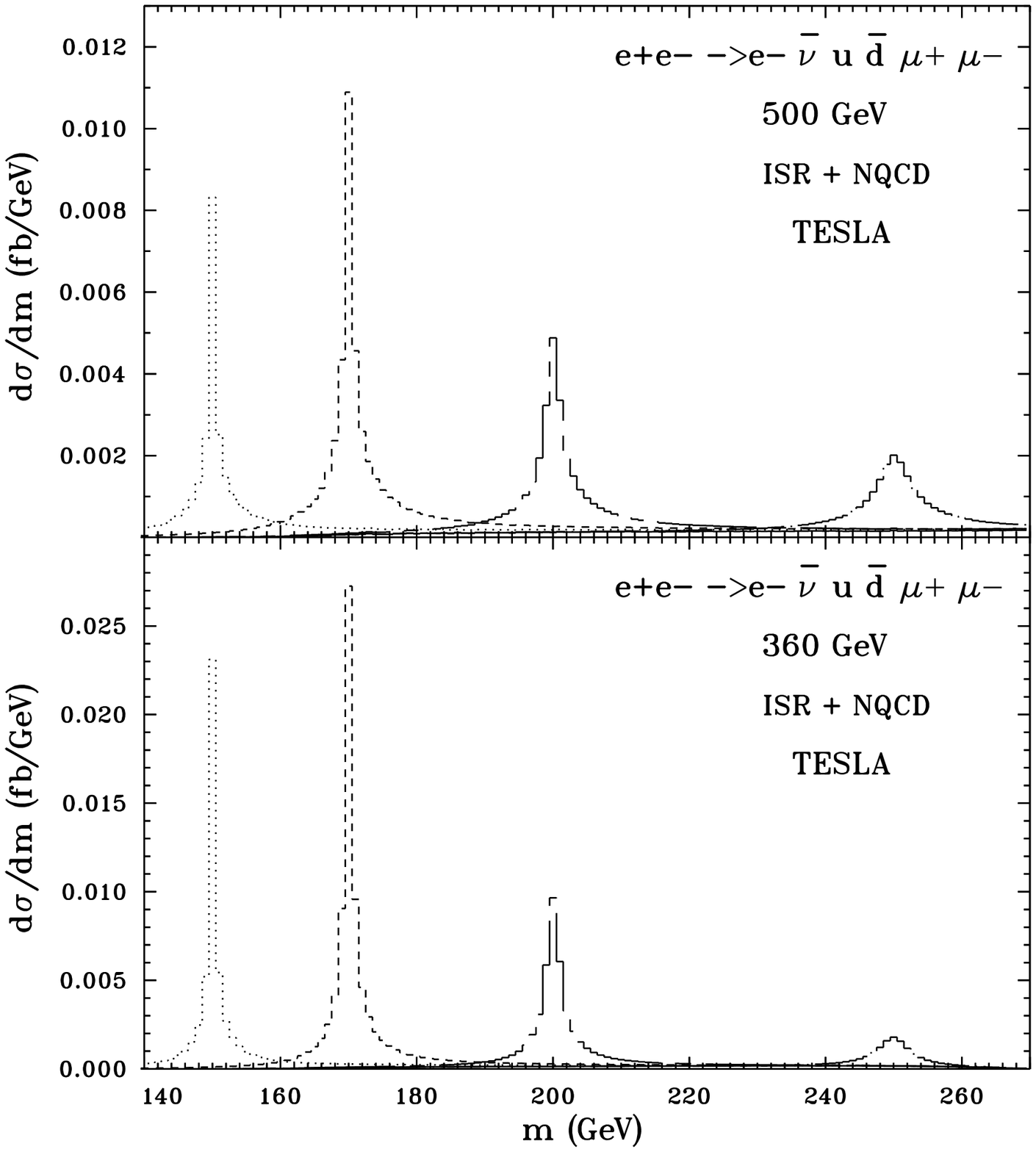}}
\end{picture}
\end{center}
\vspace{0.1cm}
\caption[]{
Reconstructed mass distributions 
The continuous line represents the total
background. The others correspond to the total cross sections for
(from left to right) $m_h$= 150, 170, 200, 250 GeV. 
Cuts: $|m(\mu^+\mu^-)-m_Z|<20$ GeV, $|m(u\bar d)-m_W|<20$ GeV.
 }
\label{h2}
\end{figure}

\begin{figure}[h]
\vspace{0.1cm}
\begin{center}
\unitlength 1cm
\begin{picture}(14.,15.)
\put(-1.8,-2.){\includegraphics{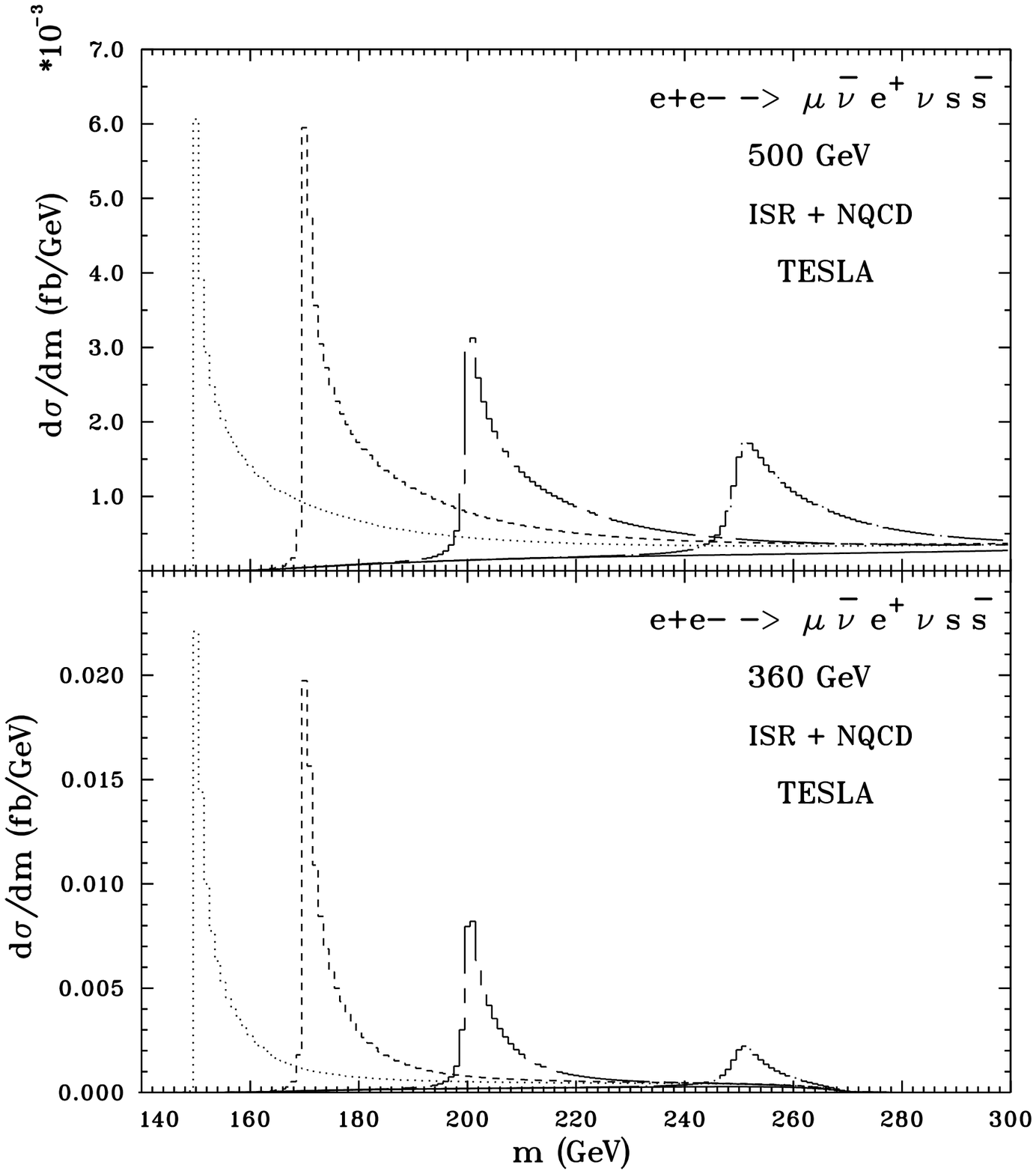}}
\end{picture}
\end{center}
\vspace{0.1cm}
\caption[]{
Missing mass distributions.
The continuous line represents the total 
background. The others correspond to the total cross sections for 
(from left to right) $m_h$= 150, 170, 200, 250 GeV.
Cuts: $|m(s\bar s)-m_Z|<20$ GeV.
}
\label{h3}
\end{figure}

\begin{figure}[h]
\vspace{0.1cm}
\begin{center}
\unitlength 1cm
\begin{picture}(14.,15.)
\put(-1.8,-2.){\includegraphics{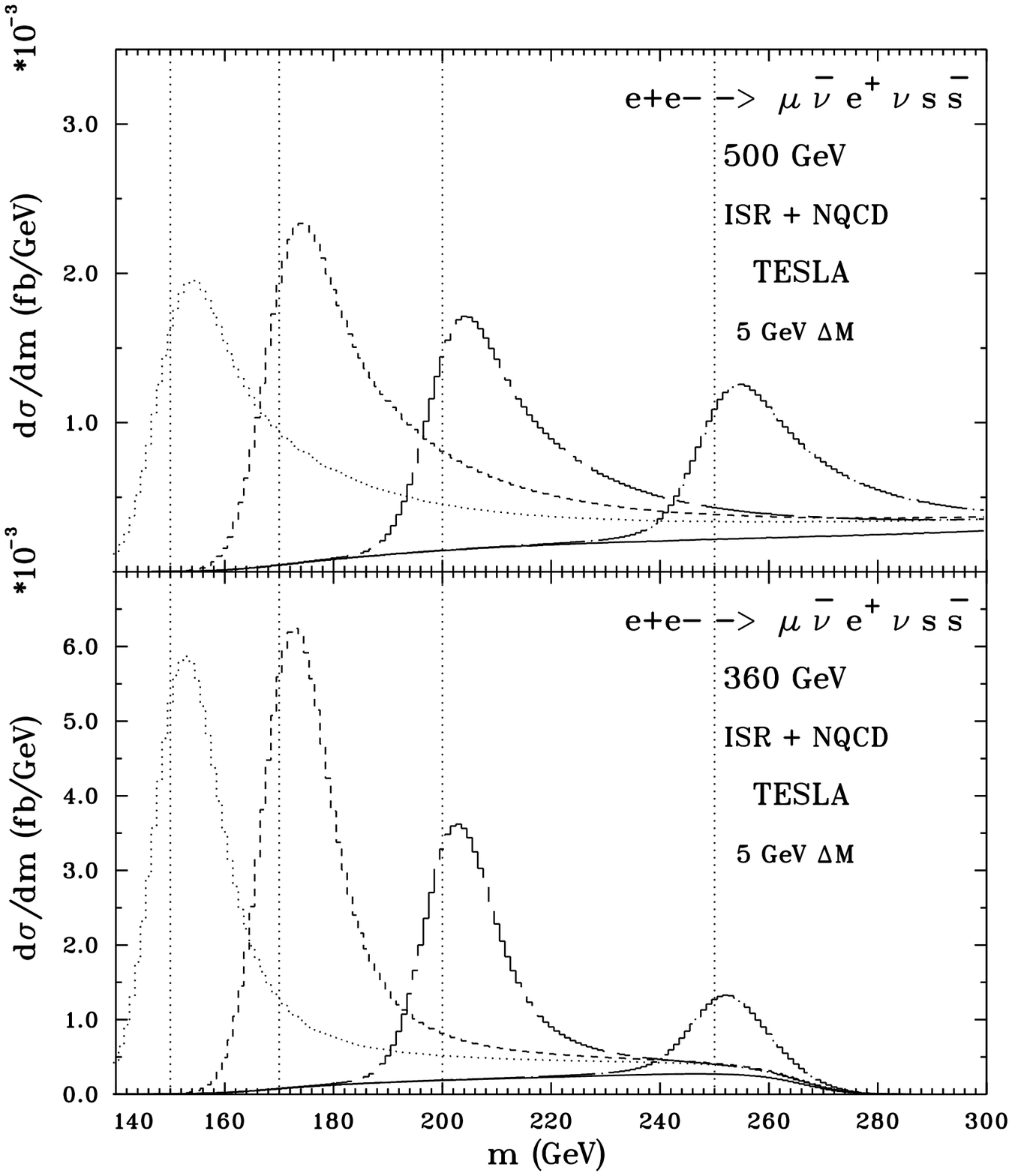}}
\end{picture}
\end{center}
\vspace{0.1cm}
\caption[]{
Missing mass distributions with gaussian smearing.
The continuous line represents the total
background. The others correspond to the total cross sections for
(from left to right) $m_h$= 150, 170, 200, 250 GeV.
Cuts: $|m(s\bar s)-m_Z|<20$ GeV.}
\label{h4}
\end{figure}

\begin{figure}[h]
\vspace{0.1cm}
\begin{center}
\unitlength 1cm
\begin{picture}(14.,15.)
\put(-1.8,-2.){\includegraphics{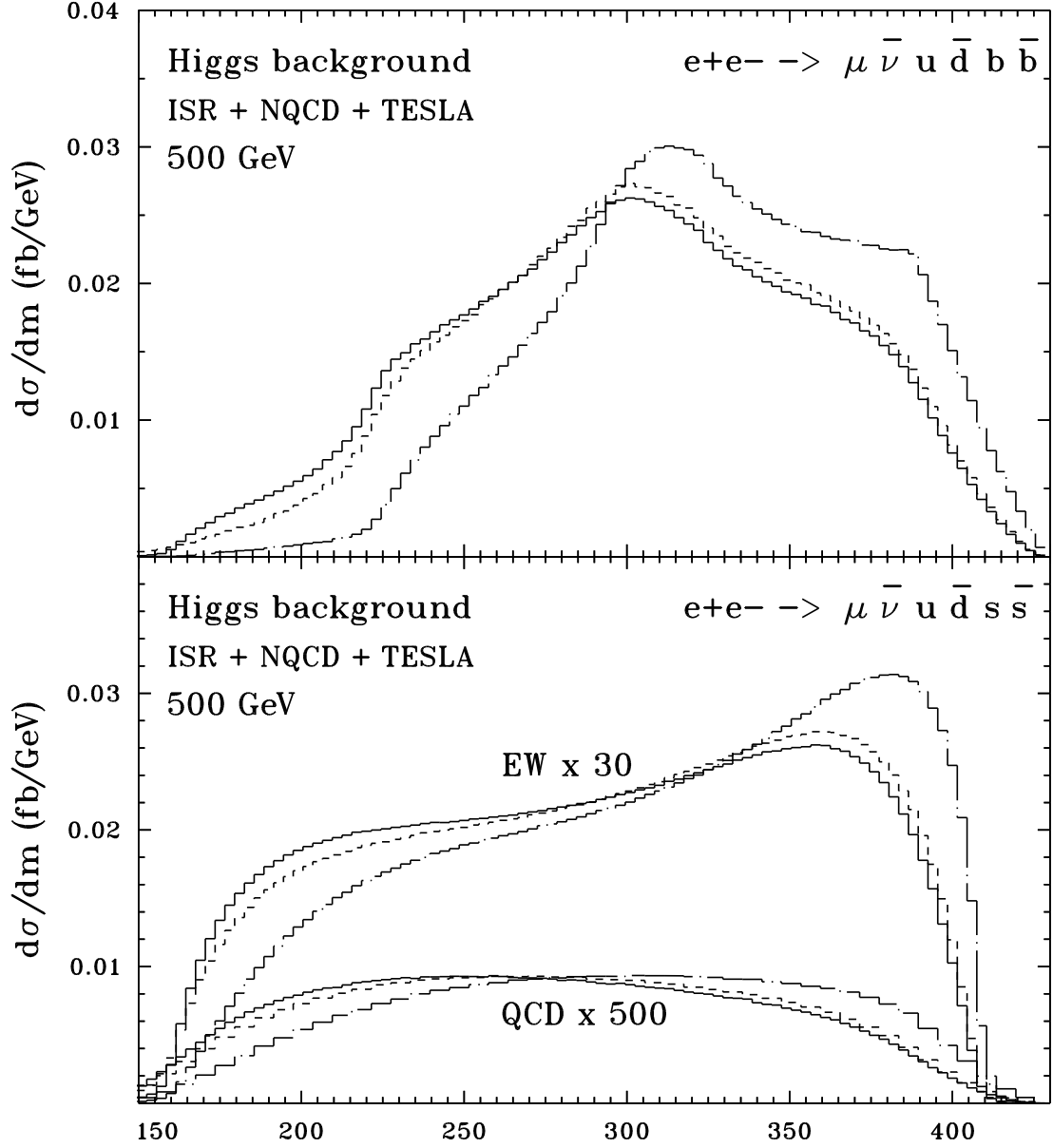}}
\end{picture}
\end{center}
\vspace{0.1cm}
\caption[]{Invariant mass distributions for higgs background. The continuous 
line  corresponds to ($\mu \bar \nu
u \bar d$)  mass, the dashed line to the {\it reconstructed} 
and the chaindot to the {\it missing} one.
Cuts:  $|m(u\bar d)-m_W|<20$ GeV, $|m(q \bar q)-m_Z|<20$ GeV ($q=b$ or $s$).
 }
\label{h5}
\end{figure}

\begin{figure}[h]
\vspace{0.1cm}
\begin{center}
\unitlength 1cm
\begin{picture}(14.,15.)
\put(-1.8,-2.){\includegraphics{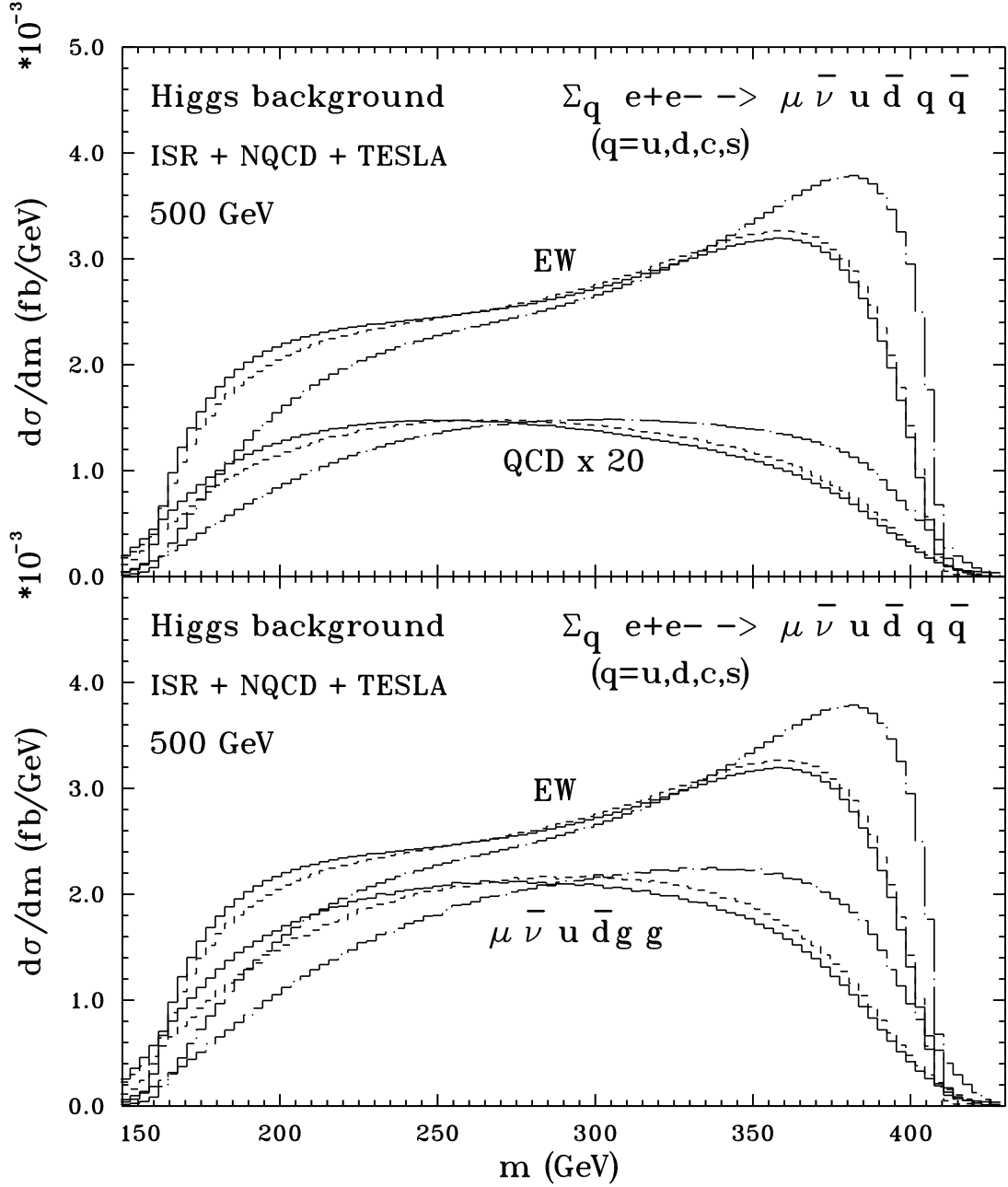}}
\end{picture}
\end{center}
\vspace{0.1cm}
\caption[]{Invariant mass distributions for higgs background. The continuous 
line  corresponds to ($\mu \bar \nu
u \bar d$)  mass, the dashed line to the {\it reconstructed} 
and the chaindot to the {\it missing} one.
Cuts: $|M_1-m_Z|<20$ GeV, $|M_2-m_W|<20$ GeV. $M_1$ and $M_2$
are the invariant masses of the pairs of strong particles which are
contemporarily nearer to $m_Z$ and $m_W$.
 }
\label{h6}
\end{figure}

\begin{figure}[h]
\vspace{0.1cm}
\begin{center}
\unitlength 1cm
\begin{picture}(14.,15.)
\put(-1.8,-2.){\includegraphics{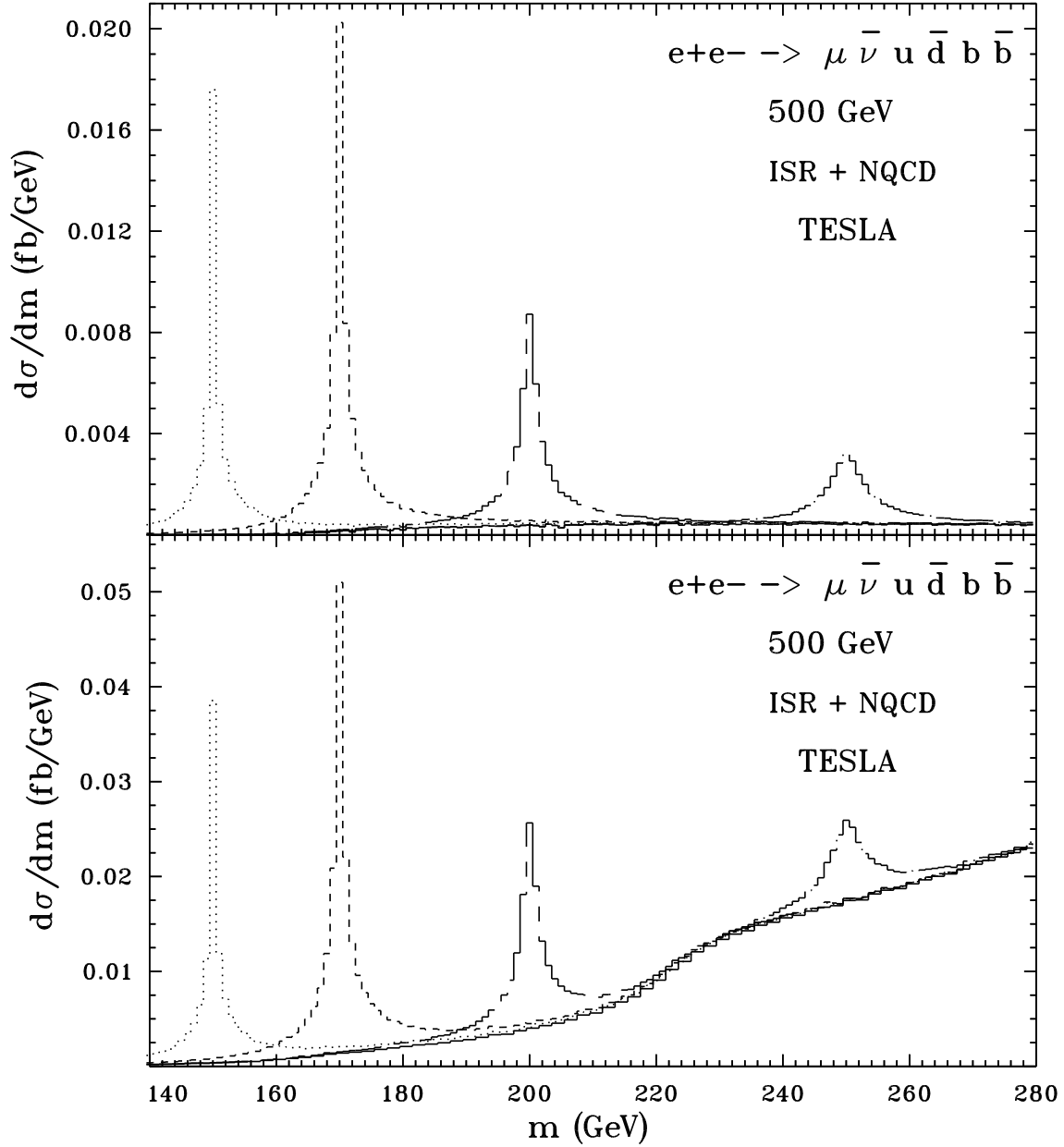}}
\end{picture}
\end{center}
\vspace{0.1cm}
\caption[]{Reconstructed mass distributions with 
$|m(b\bar b)-m_Z|$ and $|m(u\bar d)-m_W|<$ 20 GeV and, in the upper 
part only,  $|M-m_{top}|>$40 GeV ($M=m(bu\bar d)$ and 
$m(\bar bu\bar d)$). The continuous line represents the total 
background. The others correspond to the total cross sections for 
(from left to right) $m_h$= 150, 170, 200, 250 GeV.
}
\label{h7}
\end{figure}

\begin{figure}[h]
\vspace{0.1cm}
\begin{center}
\unitlength 1cm
\begin{picture}(14.,15.)
\put(-1.8,-2.){\includegraphics{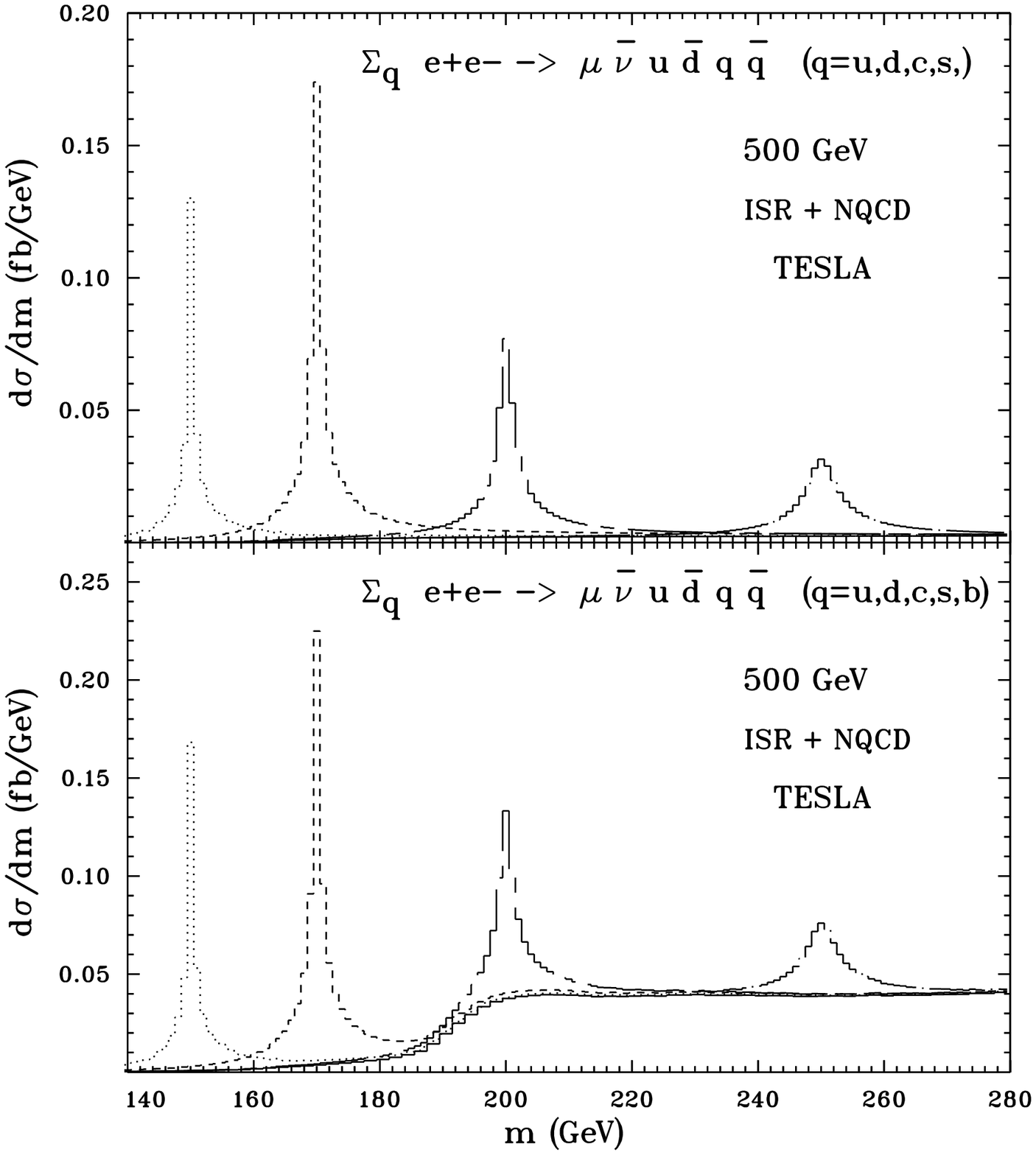}}
\end{picture}
\end{center}
\vspace{0.1cm}
\caption[]{
Reconstructed mass distributions. 
The continuous line represents the total
background. The others correspond to the total cross sections for
(from left to right) $m_h$= 150, 170, 200, 250 GeV. 
Cuts: $|m(q \bar q)-m_Z|<20$ GeV, $|m(u\bar d)-m_W|<20$ GeV.
}
\label{h8}
\end{figure}

\begin{figure}[h]
\vspace{0.1cm}
\begin{center}
\unitlength 1cm
\begin{picture}(14.,15.)
\put(-1.8,-2.){\includegraphics{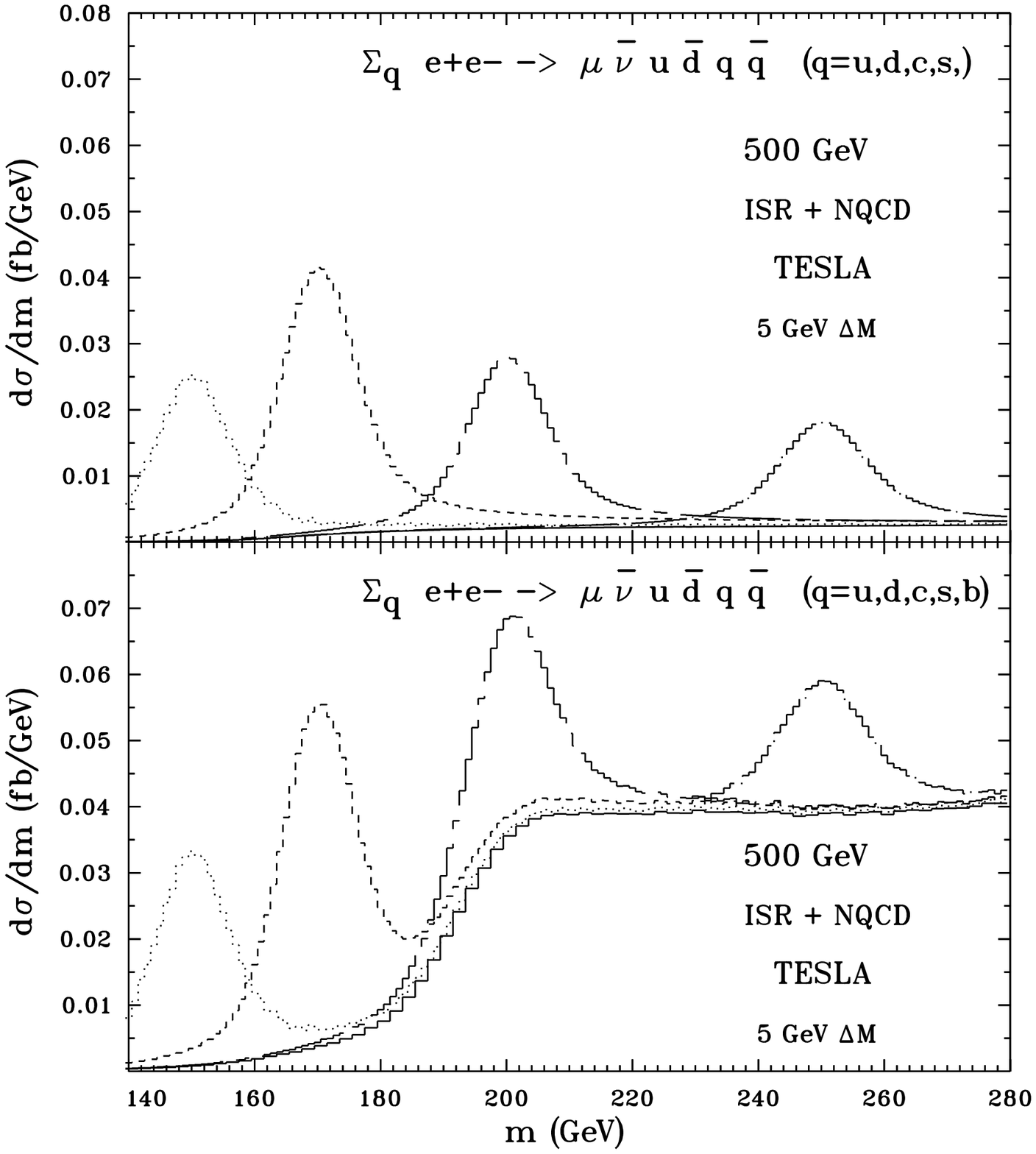}}
\end{picture}
\end{center}
\vspace{0.1cm}
\caption[]{
Reconstructed mass distributions with gaussian smearing. 
The continuous line represents the total
background. The others correspond to the total cross sections for
(from left to right) $m_h$= 150, 170, 200, 250 GeV. 
Cuts: $|m(q \bar q)-m_Z|<20$ GeV, $|m(u\bar d)-m_W|<20$ GeV.
}
\label{h9}
\end{figure}

\end{document}